\documentclass{aa}
\usepackage{url}
\usepackage{graphicx}
\usepackage{hyperref}
\usepackage{txfonts}
\usepackage{xspace}

\def\maxi{MAXI\,J1348{\ensuremath{-}}630\xspace}
\def\ero{\textit{SRG}/eROSITA\xspace}

\def\xmmn{\textit{XMM-Newton}\xspace}
\def\cxo{\textit{Chandra}\xspace}
\def\swi{\textit{Swift}/XRT\xspace}
\def\ga{\textit{Gaia}\xspace}
\def\sh{\texttt{\textit{StarHorse}}\xspace}
\def\xspec{\texttt{\textit{Xspec}}\xspace}

\begin{document}

\title{A giant X-ray dust scattering ring around the black hole transient \maxi\, discovered with \ero\thanks{Based on observations obtained with \xmmn, an ESA science mission with instruments and contributions directly funded by ESA Member States and NASA}}
\titlerunning{Giant dust ring}

\author{G.~Lamer\inst{1}\and
A.D.~Schwope\inst{1}\and
P.~Predehl\inst{2}\and
I.~Traulsen\inst{1}\and
J.~Wilms\inst{3}\and
M.~Freyberg\inst{2}
}
\institute{Leibniz-Institut f\"ur Astrophysik Potsdam (AIP), An der Sternwarte 16, 14482 Potsdam, Germany \\
\email{glamer@aip.de}
\and 
     Max-Planck-Institut f\"ur extraterrestrische Physik,
     Gie{\ss}enbachstra{\ss}e, 85748 Garching, Germany
\and
Dr.\ Karl Remeis-Sternwarte \& Erlangen Centre for Astroparticle Physics, Friedrich-Alexander-Universit\"at Erlangen-Nürnberg, Sternwartstra{\ss}e 7, 96049, Bamberg, Germany
}

\authorrunning{Lamer et al.}
\date{}


\abstract{We report the discovery of a giant dust scattering ring around the Black Hole transient \maxi with \ero\ during its first X-ray all-sky survey.
During the discovery observation in February 2020 the ring had an outer diameter of $1.3\,\mathrm{deg}$, growing to $1.6\,\mathrm{deg}$ by the time of the second all sky survey scan in August 2020. This makes the new dust ring the by far largest X-ray scattering ring observed so far. Dust scattering halos, in particular the rings found around transient sources, offer the possibility of precise distance measurements towards the original X-ray sources. We combine data from \ero,  \xmmn, MAXI , and \ga to measure the geometrical distance of \maxi. The \ga\ data place the scattering dust at a distance of 2050\,pc, from the measured time lags and the geometry of the ring, we find MAXI J1348$-$630 at a distance of 3390\,pc with a statistical uncertainty of only 1.1\% and a systematic uncertainty of 10\% caused mainly by the parallax offset of Gaia. This result makes MAXI J1348$-$630 one of the black hole transients with the best determined distances.
The new distance leads to a revised mass estimate for the black hole of 
$11\pm 2\, M_\odot$, the transition to the soft state during the outburst occurred when the bolometric luminosity of MAXI J1348$-$630 had reached $1.7\%$ of its Eddington luminosity.}

\maketitle

\section{Introduction}
Similar to optical light, X-rays of cosmic sources are affected by the interstellar medium in our Galaxy. These effects consist on the one hand of photo-electric absorption and on the other hand of dust extinction  \citep{2016MNRAS.458.1345C}, which is caused by photo absorption and by scattering from dust grains, where in contrast to visible light the scattering of X-rays takes place at small angles. The scattered radiation forms a halo around the point source, such that both components of the extinction can often be determined in a single observation. Such observations allow us to draw conclusions about the physical and chemical properties of the interstellar dust \citep[][and references therein]{mauche_gorenstein86, mathis_lee91, predehl_schmitt95, draine03, 2011ApJ...738...78X, 2017ApJ...839...76C}.

Since the scattered light has to travel a longer distance than the direct light, brightness variations of the central source appear with a delay in the ``echo'' of the dust scattering halo. This was proposed as early as 1973 as a method to determine the geometrical distance of X-ray sources  \citep{truemper_schoenfelder73}, but was only realised 27 years later through a \cxo\ observation of Cyg X-3 \citep{predehl+00}. As a rule of thumb, the delay of the echo for a source at a distance of 5\,kpc and dust in the middle between source and observer for sake of simplicity is a few hours at $1'$, some weeks at $10'$, and one year at half a degree. 

Since in practice most X-ray sources show variability on all timescales, the variability of the halo is very complex as it is the convolution of the impulse response of the halo with the source variability. This makes distance measurements with dust scattering halos challenging. The best sources for distance measurements with scattering halos are, therefore, transient X-ray sources, such as X-ray binaries (XRB), soft gamma repeaters (SGRs), or gamma-ray bursts (GRBs) as for these well-defined light echos in the form of distinct rings of X-rays will occur that grow with time. In the case of sources whose distance is known (or very large), such an observation allows a tomography of the dust distribution. Utilising an \xmmn\ observation of expanding rings around GRB031203, \citet{vaughan+04} were the first to succeed with the determination of the distance of two dust clouds at 880\,pc and 1.3\,kpc. \cite{clark04}, with a \cxo\ observation of the X-ray pulsar 4U1538$-$52, was able to determine both its distance (4.5\,kpc) and that of three layers of dust in between (1.3\,kpc, 2.56\,kpc, and 4.05\,pc). \cite{2015ApJ...806..265H} managed to identify a total of four rings (``Lord of the Rings'') around Cir X-1. A general consideration of the possibilities of such observations can be found in \cite{corrales+19}. 
For dust scattering rings where either the distance of the source or of the scattering dust is known from other measurements the second distance can be calculated geometrically using the ring radius and the time lag of the scattered X-rays.
In other cases both distances have been constrained by modelling the temporal intensity evolution of the expanding rings \citep[e.g.][]{2010ApJ...710..227T}. However, this method requires the knowledge of the dust scattering cross section and the results therefore depend on the choice of the model for the dust composition and grain size distribution.

\begin{figure*}[t]
\resizebox{0.49\hsize}{!}{\includegraphics[clip=]{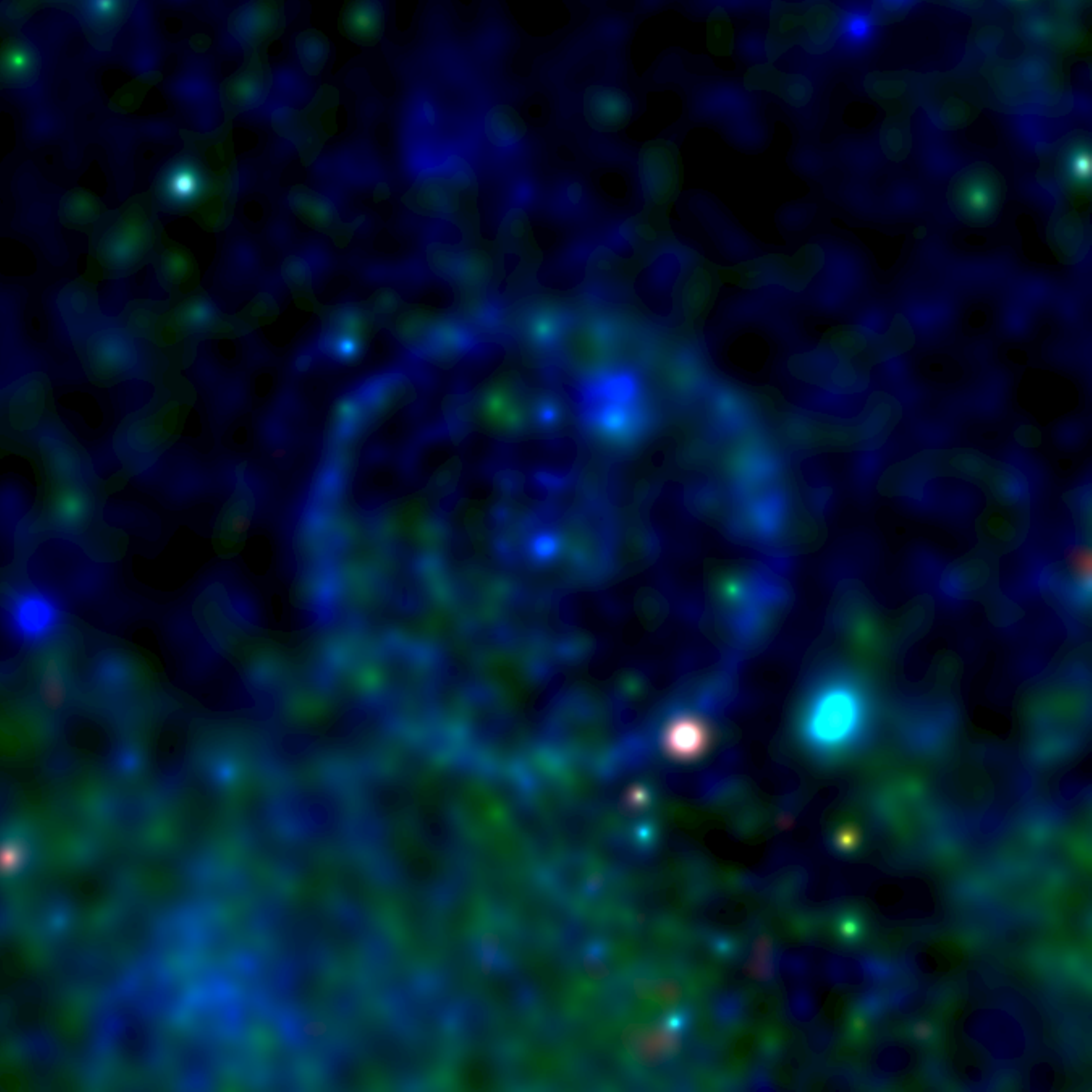}}\hfill
\resizebox{0.49\hsize}{!}{\includegraphics[clip=]{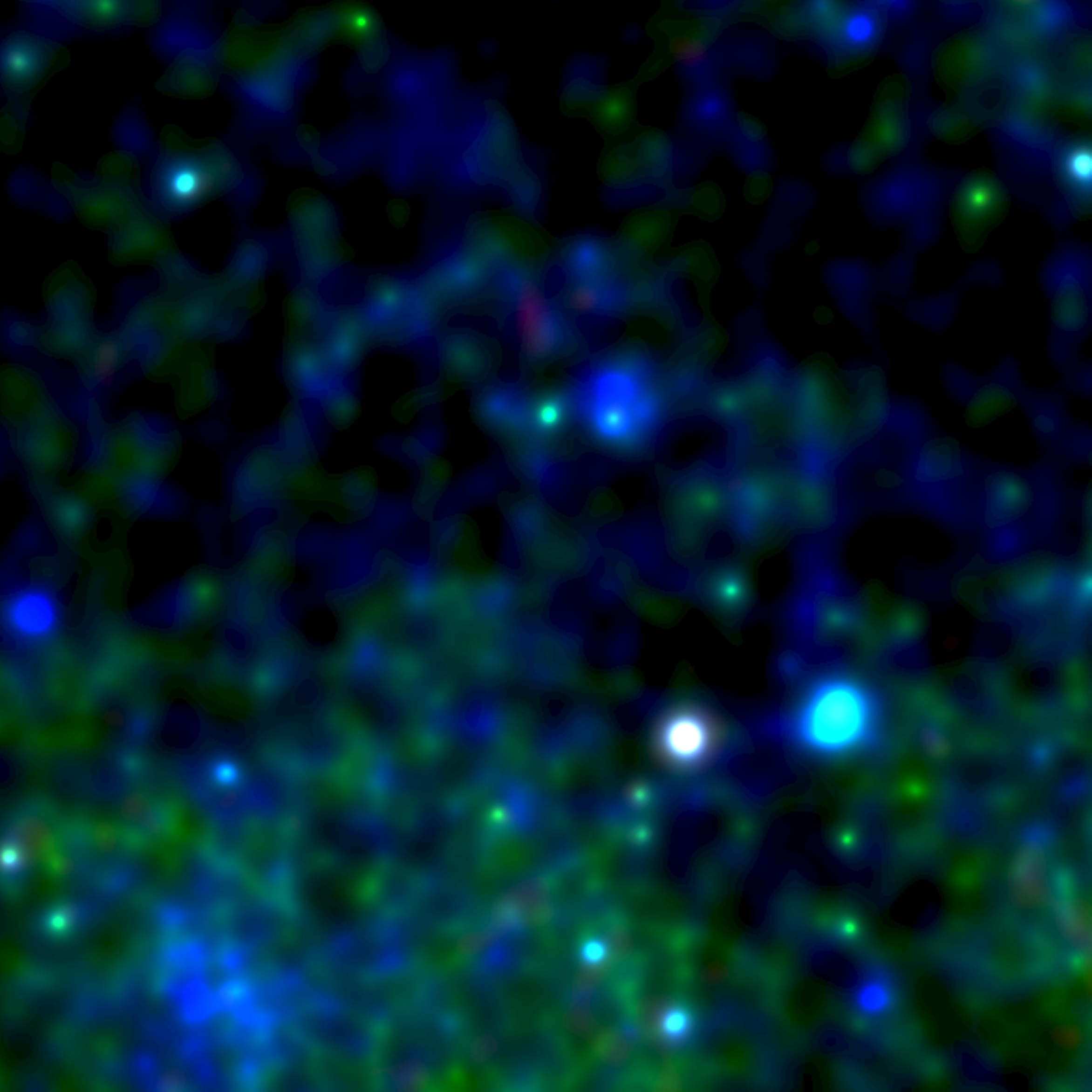}}
\caption{False-colour images of the eRASS1 (18-24 Feb 2020, {\it left}) and  eRASS2 (17-22 Aug 2020, {\it right})  observations in the energy bands $0.2-0.6$ keV (red), $0.6-1.0$ keV (green), $1.0-2.3$ keV (blue), adaptively smoothed. The size of the images is $3\degr \times 3\degr$. North is at the top and East to the left.
\label{f:e1_rgb}}
\end{figure*}

Following the initial report of the detection on 2019 January 26, 03:16, of a new, bright X-ray transient with MAXI/GSC onboard the International Space Station \citep{2019ATel12425....1Y}, comprehensive follow-up activities were initiated \citep[e.g.][]{2019ATel12447....1S, 2019ATel12456....1R, 2019ATel12456....1R, 2019ATel12441....1A, 2019ATel12470....1C}. The monitoring observations of \maxi\ with MAXI, follow-up X-ray observations with NICER, \swi, \textit{INTEGRAL} and \textit{INSIGHT}, the identification of an optical counterpart and its spectral shape, and  radio observations classified the object as a likely Black Hole Transient (BHT) at $\alpha_\mathrm{J2000}=13^\mathrm{h}48^\mathrm{m}12\fs73$, $\delta_\mathrm{J2000}=-63^\circ 16' 26\farcs8$ \citep{2019ATel12434....1K} ($b^\mathrm{II} = 309\fdg26414$, $l^\mathrm{II} = -1\fdg10302$). The maximum flux of the source at $1.0\times10^{-7}\,\mathrm{erg}\,\mathrm{s}^{-1}\,\mathrm{cm}^{-2}$ (2--20\,keV, \,$\sim 4\,$Crab) was reached about 2\,weeks after the outburst, followed by a fast decrease of brightness.

Comprehensive analyses of the MAXI and \swi data assembled during the outburst  of the BHT and the following months are presented in \cite{2020ApJ...899L..20T} and  \cite{2020ApJ...897....3J}. Hardness intensity diagrams show the spectral evolution of the source to follow the typical track of BHTs and lead to a distance estimate of 3--4\,kpc, likely in front of the Scutum-Centaurus arm.

During its first all sky survey, \ero\ scanned the sky area of \maxi\ from 2020 Feb 18 to 2020 Feb 24. During routine inspection of the data products generated after daily ground contacts, a large ($>$1\degr) and almost perfect circular ring was recognised. A central source was also found coincident with \maxi. The structure is thus naturally associated with \maxi and interpreted to be caused by scattered X-rays from the initial burst. Further inspection of the X-ray image revealed further candidate rings (or partial rings) in- and outside of the main ring. An \xmmn\ follow-up observation in Directors Discretionary Time (DDT) was immediately triggered with the aim to confirm the initial findings, to better qualify the point source contamination of the structure, and to facilitate the spatial and spectral analysis of the rings through improved photon statistics. 

In this paper, we present the results of these observations, concentrating on the scattering halo. Compared to earlier dust scattering halo measurements, which had in common that discovery and follow-up was possible with just one of the contemporary wide-angle X-ray cameras onboard, e.g., \xmmn, \cxo, or \swi, the case of the dust scattering echo around \maxi\ is different. With a size of more than 1\degr\ in diameter such a structure can only be discovered through scanning observations similar to those that \ero\ has been doing since 2019 December 12. With its large field of view of $62'$ and its large light collecting power (comparable to \xmmn) the instrument is perfectly suited to make such discoveries, provided the relevant time scales fit with the geometrical set-up of the emitter and the scatterer.
The remainder of this paper is structured as follows. In Sect.~\ref{sec:obs} we describe the eROSITA and \xmmn observations of \maxi. We then discuss the scattering halo, determine the distance to the scatterer and \maxi  and perform a spectral analysis of the scattering ring and the post-outburst transient in Sect.~\ref{s:ana}, and summarise our results in Sect.~\ref{sec:discuss}.

\section{X-ray observations}\label{sec:obs}

\subsection{eROSITA observations}

Launched on 2019 July 13 into an orbit around the $L_2$ point of the Earth-Sun system, the eROSITA instrument on board the Spectrum-X-Gamma spacecraft (SRG, Sunyaev et al. 2021, in prep.) consists of seven  X-ray camera assemblies behind seven identical and co-aligned Wolter telescopes. See \citet{merloni+12} and \citet{Predehl+20} for a description of the instrument and its science goals. Sensitive in the 0.2--8\,keV band and with a peak on-axis effective area of over $2000\,\mathrm{cm}^2$, since the end of its performance verification phase on 2019 December 13, until the end of 2023 eROSITA will perform the deepest X-ray all sky survey to date. The eROSITA survey is a slew survey. The telescopes scan along great circles that are approximately perpendicular to the ecliptic, with a rotational axis pointing towards the Earth. This way the whole sky is scanned within 6\,months. The rotational period of 4\,h together with the $1^\circ$ field of view means that any patch of sky is seen every 4\,h for several eROSITA slews (depending on the ecliptical latitude of the object), and then again half a year later.

\subsubsection{eRASS1}
\label{s:erass1}

The area around \maxi\ was scanned 31 times with \ero\ between MJD 58897.825 and MJD 58903.344. We assume a midpoint of the observations of MJD 58900.583 (2020-02-21 14:00:00 UTC), 391.3\,d after the MAXI discovery of the burst.  The dust scattering ring, which is rather inconspicuous in unsmoothed event images, was first discovered in smoothed sky maps produced from these data.
The vignetted exposure, i.e., the equivalent exposure time of an on-axis observation with all 7 telescopes, varies between 150\,s and 300\,s over the area of the scattering ring.  We have created images in the energy bands 0.2--0.6\,keV, 0.6--1.0\,keV and 1.0--2.3\,keV. 
For the purpose of visualisation the 3 energy band images were exposure corrected and adaptively smoothed using the eSASS (Brunner et al. 2021, in prep.)  task {\tt erbackmap} and
combined  into a pseudo RGB image (Fig.~\ref{f:e1_rgb}). 

\begin{figure}
\resizebox{\hsize}{!}{\includegraphics[trim=1.4cm 1cm 2cm 1cm,clip=true]{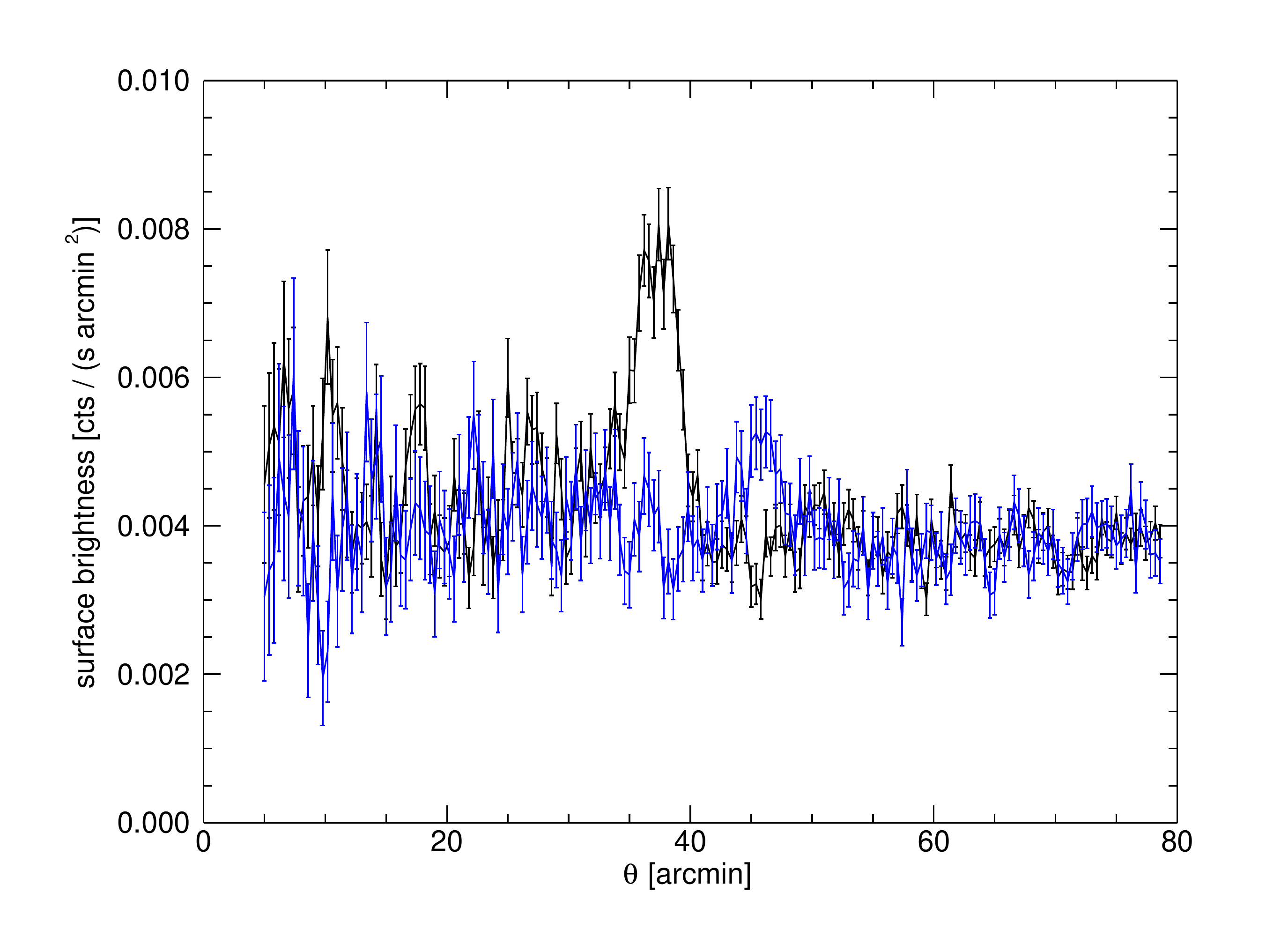}}
\caption{Radial profile of the eRASS1 (black) and eRASS2 (blue) images in the energy band 1.0-2.3 keV with a resolution of  20 arcsec. The scattered emission is  detected at radii $\sim (34 - 40)$\,arcmin in eRASS1 and between $\sim (40 - 47)$\,arcmin in eRASS2.
   }
    \label{f:e01prof}
\end{figure}

From the unsmoothed, exposure corrected image in the 1.0--2.3\,keV band we derived a radial profile of the surface brightness centered at the position of \maxi\ (Fig.~\ref{f:e01prof}). The scattered radiation is detected in an annulus with inner and outer radii of $34'$ and $40'$, respectively.
 
\subsubsection{eRASS2}
The sky area of the scattering ring was covered again during the second eROSITA all-aky survey (eRASS2) with 31 single scans between MJD 59078.597 and MJD 59083.772, the  midpoint is MJD 59081.201 (2020-08-20 04:50:00 UTC), or 571.9\,days after the  MAXI burst detection. The vignetted exposure of the relevant sky region varies only slightly between 203 s and 210 s.

The data were processed in the same way as the eRASS1 data. The dust echo is still visible at the now expected radius, about 1.2 larger than during the eRASS1 observation but with significantly lower surface brightness (Figs.~\ref{f:e1_rgb} and \ref{f:e01prof}).

\subsection{\xmmn}

\begin{figure*}[t]
\includegraphics[width=0.495\linewidth]{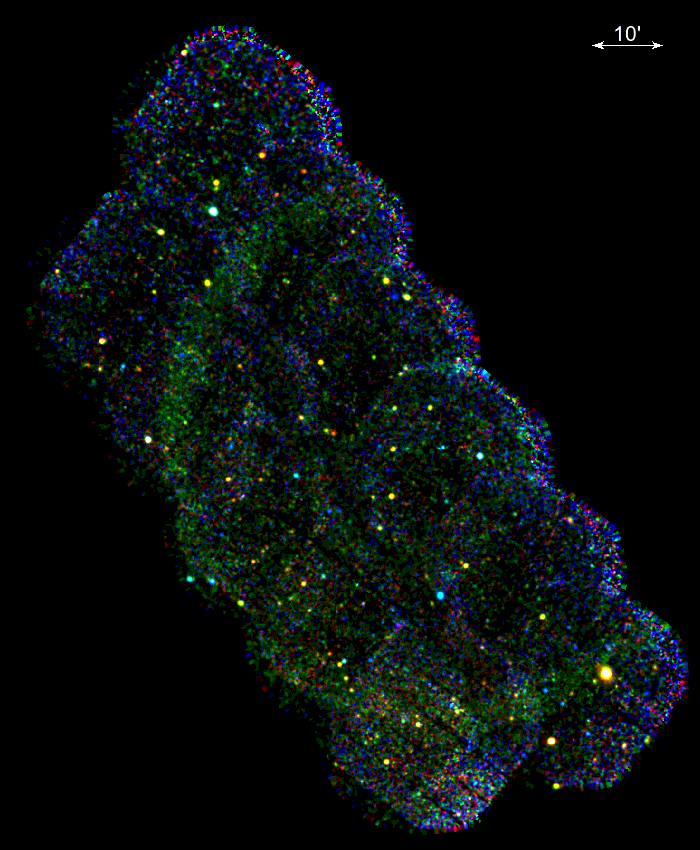}\hfill%
\includegraphics[width=0.495\linewidth]{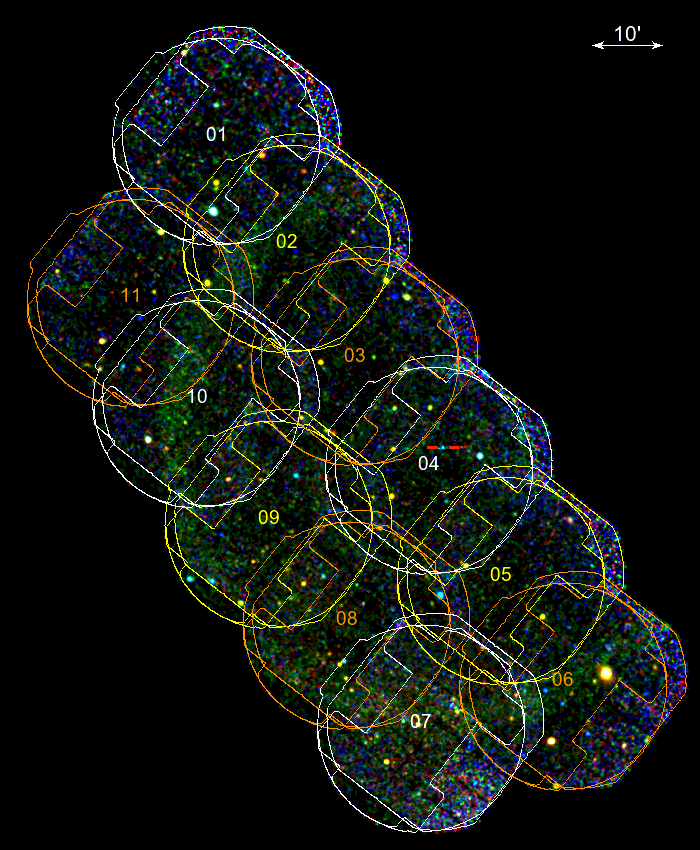}
\caption{False-colour image of the XMM-Newton mosaic-mode observations in the energy bands 0.2--0.8\,keV (red), 0.8--2.2\,keV (green), 2.2--7.0\,keV (blue), adaptively smoothed with a 2-pixel tophat. The right panel illustrates the eleven sub-pointings with the contours of the three detectors EPIC/pn, MOS1, and MOS2. The position of \maxi is marked near the centre of sub-pointing 04.
     \label{fig:epicmosaic}}
\end{figure*}

MAXI\,J1348$-$630 was observed on 2020 March 10 and 11 by \xmmn in Directors Discretionary Time in mosaic mode (observation identifier 0870590101). EPIC/pn, MOS1, and MOS2 were operated in full-frame mode using the medium filter.  Eleven EPIC pointings with a total exposure time of 75\,ks covered \maxi and almost 60\% of the dust ring. The first pointing was exposed for 5.0\,ks in MOS and 2.5\,ks in pn, the last pointing for 6.2\,ks in MOS and 5.9\,ks in pn, and the other pointings for 6.6\,ks in all instruments. Figure~\ref{fig:epicmosaic} shows the contours of the EPIC/pn and MOS detectors for each sub-pointing. The mosaic-mode data were reduced using the \textit{XMM-Newton} Science Analysis System \citep[SAS,][]{2004ASPC..314..759G} and split into the sub-pointings by the task \texttt{emosaic\_prep}.

To create mosaic images and to perform source detection, they were projected onto common coordinates centered at the mean attitude of all pointings by the task \texttt{edetect\_stack} \citep{2019A&A...624A..77T}. Stacked source detection by \texttt{edetect\_stack} was used to mask sources in the event lists, from which the radial profile and the spectra were derived. Since images with spatially very inhomogeneous background are prone to misinterpretation of background features as extended sources, the task parameters were optimised for point sources. \texttt{esplinemap} in smoothing mode was run with the parameters given by \citet{2019A&A...624A..77T}, and the source-detection task \texttt{emldetect} was run without extent fitting. In addition to \maxi, 280 sources were found. Circular regions with brightness-dependent radius were cut out, a mask including the good source-free regions generated, and source-excised event lists created for each pointing and instrument.

Images of all pointings and instruments were created in common coordinates using the task \texttt{edetect\_stack} in the five standard XMM-Newton energy bands (1) 0.2--0.5\,keV, (2) 0.5--1.0\,keV, (3) 1.0--2.0\,keV, (4) 2.0--4.5\,keV, (5) 4.5--12.0\,keV, and in three energy bands optimised for the flux maximum of the dust ring, derived from the spectra (Sect.~\ref{s:ringspec}): 0.2--0.8\,keV, 0.8--2.2\,keV, and 2.2--7.0\,keV. The individual images were corrected for background emission and different exposure times and combined into mosaics as follows.

The background of the dust-ring observations is composed of the instrumental background, the local particle background in the orbit, the cosmic X-ray background, and the Galactic components along the line of sight. We model this complex mixture based on source- and ring-free regions in the sub-pointings by the task \texttt{esplinemap} in smoothing mode. For each energy band and instrument, the smoothed background maps of ten sub-pointings were averaged, scaled to the exposure of each sub-pointing and subtracted from the original images. Pointing 7 was excluded from the averaging because of its high background emission. For EPIC/pn, the out-of-time events were modelled within \texttt{esplinemap} and also subtracted. The background-subtracted images were exposure-corrected with a combination of exposure maps with and without taking the effects of the mirror vignetting into account. Division of the images by the vignetted exposure maps would give a high weight to the low-exposed outer CCD areas which would appear too bright. Therefore, we used an empirically chosen weighting factor of 0.7 for the vignetted and 0.3 for the unvignetted map. The exposure-corrected images were combined into mosaics. The mosaic image including the dust-ring structure and all sources is shown in Fig.~\ref{fig:epicmosaic}.
A radial brightness profile with 12 arcsec spacing was derived from the $1-2$\,keV mosaic image as described in Sect.~\ref{s:erass1} and shown in Fig. \ref{f:erass1_fit} (right).

Spectra were generated from the data taken during individual pointings in their genuine coordinates by the task \texttt{especget}, using the source-excised event lists. The pn and MOS spectra of \maxi were taken from a circular extraction region with a radius of 20\arcsec\ and the background from a nearby half annulus, both centered at the source position. They were grouped to include at least one count in each bin. For the ring spectra, an annular region with an inner radius of 34.5\arcmin\ and an outer radius of 41.0\arcmin\ was chosen from the radial profile (Fig.~\ref{f:erass1_fit}). Background spectra were generated from large circular source-free regions outside the ring structure and applied to all sub-pointings for which a similar background level can be expected. The background spectra generated from pointing 1 were used for the ring spectra of pointing 2 and the background of pointing 6 for 6, 8, and 9. In the pointings 7, 10, and 11, the background spectra could be used directly. The EPIC/MOS1 and MOS2 spectra of each pointing and their responses were merged by the task \texttt{epicspeccombine}. All spectra were binned to a minimum signal-to-noise ratio of 1.0 and analysed jointly using \xspec version v12.11.1 (heasoft-6.28).

\subsection{MAXI}

\begin{figure}
\resizebox{\hsize}{!}{\includegraphics[angle=270,clip=]{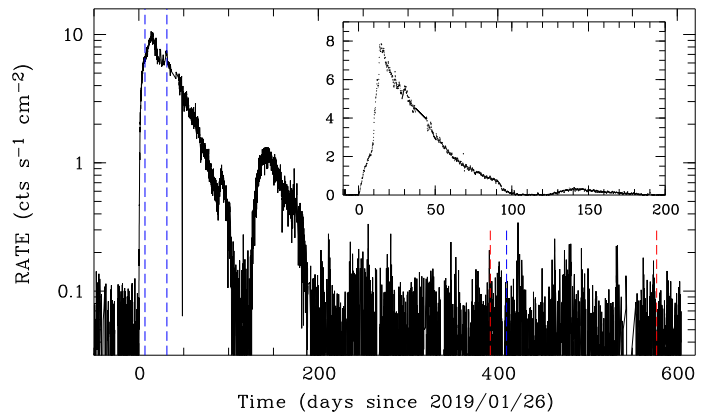}}
\caption{MAXI light curve in the 2--20\,keV band. Dashed blue lines indicate \xmmn\ observations (0831000101: 26 ksec, 0831000301: 15 ksec, 0870590101: 77 ksec), dashed red lines eRASS1 and eRASS2 observations. The inset shows the MAXI light curve in  the 2--3\,keV band in linear representation, binned in 0.1\,d intervals and interpolated in periods of missing data.}
\label{f:maxi_lc}
\end{figure}

X-ray flux at energies 2--50\,keV from \maxi was detected by MAXI for most of the period within 175 days after the initial discovery on 2019 January 26 \citep{2020ApJ...899L..20T}.
The first outburst peaked 14\,days after discovery and after a roughly exponential decay the source disappeared after 104\,days. A second, spectrally harder outburst was recorded between days 126 and 175 after discovery. Another re-brightening of \maxi\ in 2020 February was reported by \cite{2020ATel13459....1S}. The flux levels of this re-brightening were more than two orders of magnitude lower than the primary outburst and therefore this re-brightening is not relevant for the observation of dust scattering.

The dust scattering ring is detected by \ero\ and \xmmn\  at  photon energies 0.8--2\,keV, an energy band not covered by MAXI. In order to obtain a light curve in a band which matches the photon energies relevant for the scattering as closely as possible, we downloaded the 2019 MAXI data  from the HEASARC mirror. We then used the HEASOFT task {\tt mxproduct} with standard settings to extract a light curve for \maxi\ in the energy band 2--3\,keV. We removed bad stretches of data during which the line of sight to \maxi\ was blocked by the Crew Dragon space craft docked to the ISS. The remaining data were binned to a resolution of 0.1\,days and stretches of missing data were filled by means of interpolation. This light curve (inset in Fig.~\ref{f:maxi_lc}) was used as reference for the analysis of time lags between the direct flux from \maxi and the  scattered X-rays detected  by \ero\ and \xmmn.

\begin{figure}
\resizebox{\hsize}{!}{\includegraphics[width=0.49\textwidth]{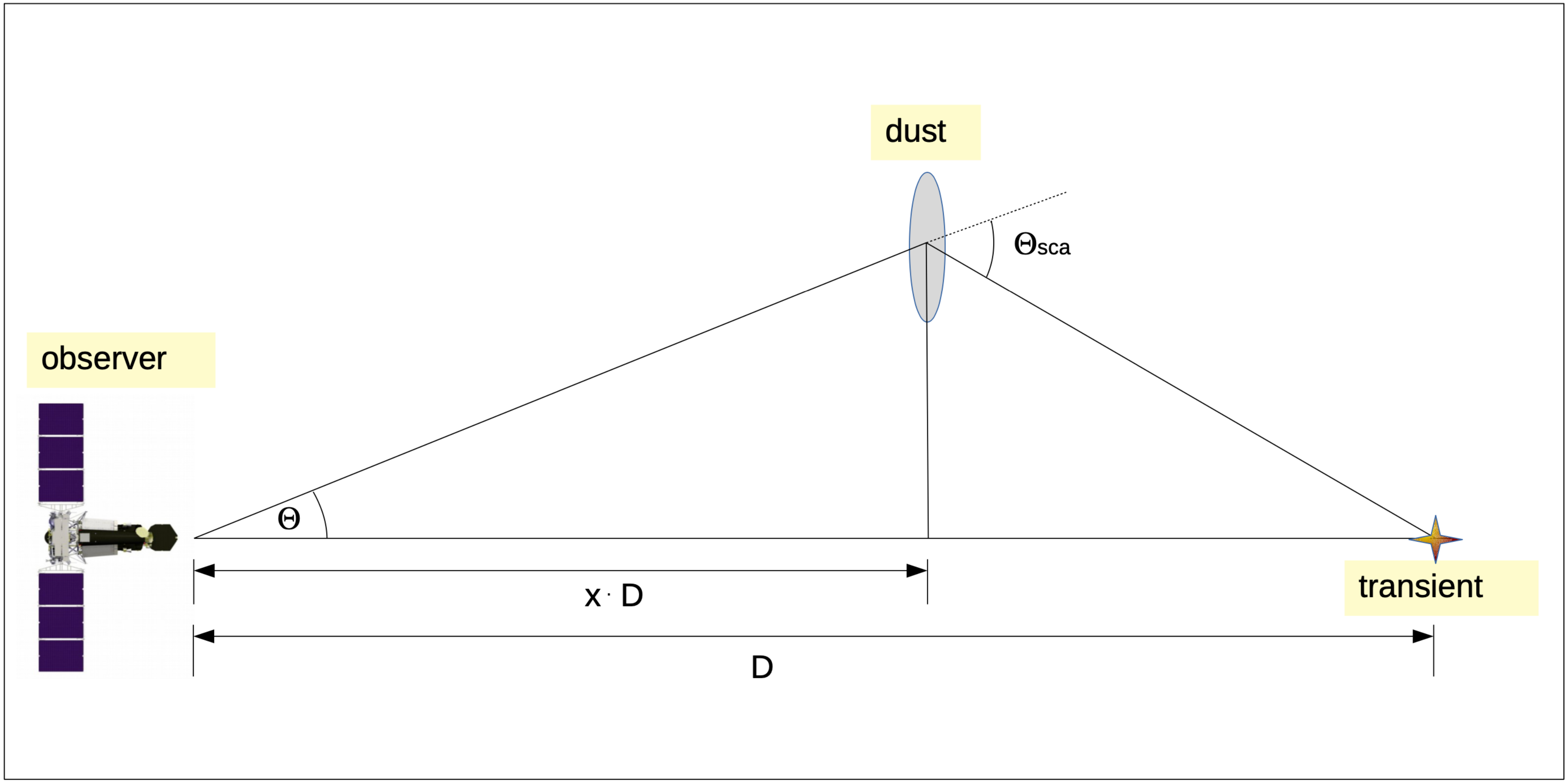}}
    \caption{Schematic drawing of the scattering geometry. The distance  to the X-ray source is denoted by D, the distance between observer and dust layer is $x \cdot D$.
The opening angle of the scattering ring is $\theta$, the X-rays are scattered by $\theta_{sca}$. }
    \label{f:sketch}
\end{figure}

\section{Analysis and results}
\label{s:ana}
\subsection{Scattering geometry}

The scattering geometry 
is illustrated in Fig.~\ref{f:sketch}. It is determined by the distance $D$ to the source, the distance to the scattering layer $xD$, the opening angle of the scattering ring, $\theta$, and the time delay $\Delta t$ between the arrival times of the original signal and the light echo at the observer. The scattered signal will arrive with a time delay of
\begin{equation}
  \label{eq:deltat}
  \Delta t = \frac{\Delta D}{c} = \frac{xD\theta^2}{2c(1-x)}
\end{equation}
where the small-angle approximation was used and where $\Delta D$ is the additional distance the X-rays take due to scattering.

Solving Eq.~\ref{eq:deltat} for the angular offset from the source position, $\theta$, results in
\begin{equation}
  \label{eq:theta}
  \theta = \sqrt{\frac{2c (1-x)}{xD} \Delta t }
\end{equation}
For well defined values of $\Delta t$ and $xD$, i.e., a sufficiently short burst of radiation and a single layer of dust at distance $xD$, a scattering ring can be observed.

Observationally, one needs to determine $\theta$ and $\Delta t$ to fix the relative geometry. If either $xD$ or $D$ can be determined independently, the absolute size of the triangle can be derived. Fortunately this is possible for the light echo around \maxi.
\

\subsection{Location of scattering dust }

Following  Fig.~\ref{f:sketch} and Eq.~\ref{eq:deltat} the distance $D$ towards the source can be determined from the time lag,  $\Delta t$, if the distance towards the scattering material $xD$ is known. In the real world, however, this measurement is complicated by the structure of the interstellar medium along the line of sight. The intensity distribution of the halo is therefore given by a convolution of time lags introduced by the dust distribution along the line of sight with the variability of the source.

\begin{figure}
\resizebox{\hsize}{!}{\includegraphics[trim=0.2cm 0.1cm 0.5cm 0.2cm,clip=true,width=0.49\textwidth]  {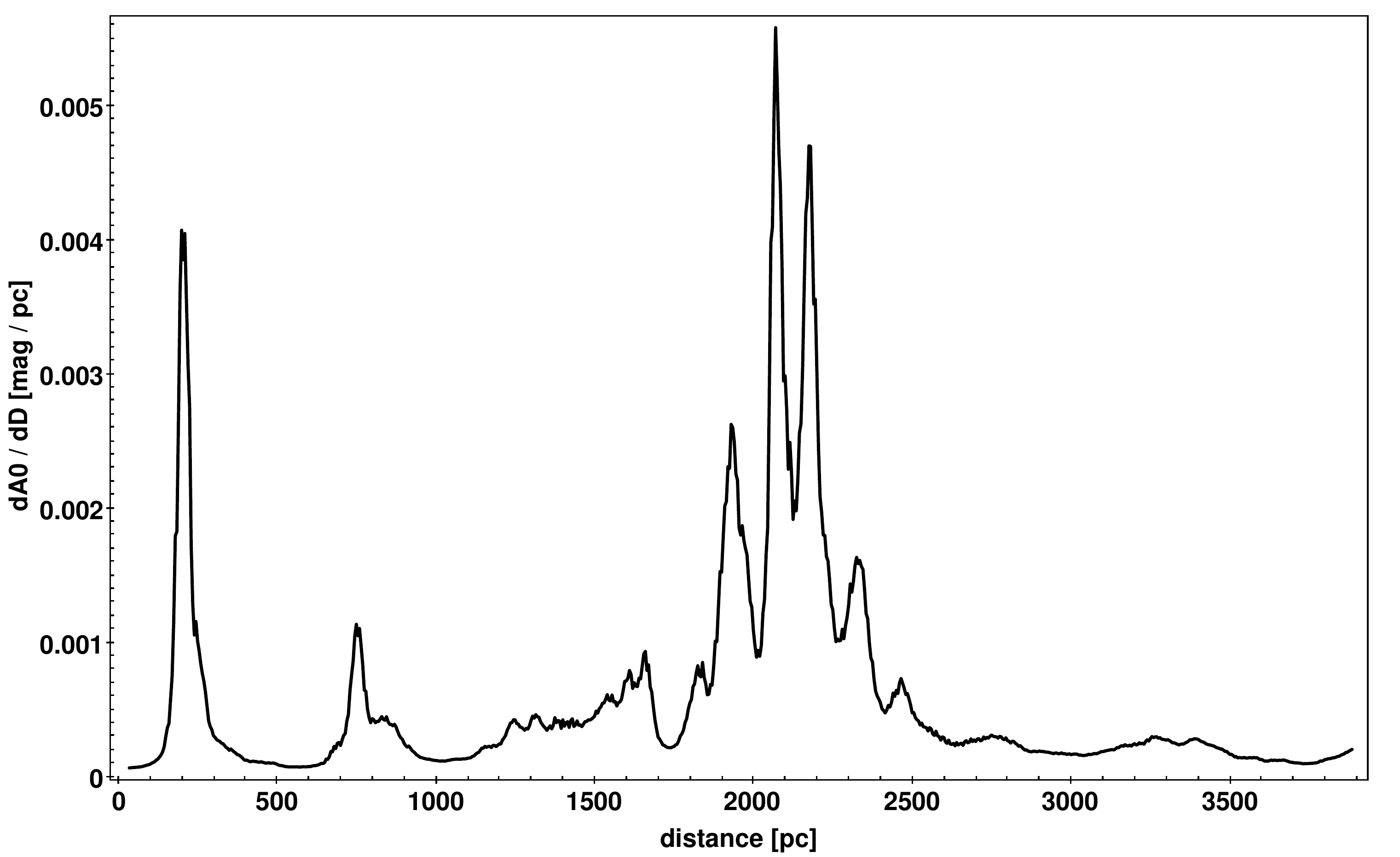}}
    \caption{ Differential extinction on the line of sight towards MAXI\,J1348$-$630 extracted from the 3D map compiled by \cite{2019A&A...625A.135L}.}  \label{f:dust_los}
\end{figure}

The circumstances leading to the giant scattering ring around \maxi\ make this event an ideal opportunity to measure the distance of \maxi. The well-defined annular shape of the dust echo is due to the basically single-peaked burst and also implies the presence of a single, well defined layer of scattering material. We can confirm this conjecture with a more detailed study of the dust distribution using new \ga data. Since the \ga data release DR2 \citep{2017A&A...599A..50A} several 3-dimensional maps of the interstellar dust extinction in our Milky Way have been published \citep[e.g.,][]{2019A&A...625A.135L,2019MNRAS.483.4277C}. The 3D dust extinction cube  published by \citet{2019A&A...625A.135L} covers a volume of $6\times 6\times 0.8$\,kpc in the solar neighbourhood with 5\,pc spatial binning and was derived by combining photometric and astrometric data from \ga\ and 2MASS for 27 million stars with \ga parallax uncertainties $ < 20 \%$. We have analysed the data cube  at the celestial position of \maxi\, where it extends to a maximum distance of ${\sim}3.9$\,kpc. The most significant regions of extinction are found at distances from the Sun between 1800 and 2200\,pc with an integrated $A_\mathrm{V} = 0.9$\,mag (Fig.~\ref{f:dust_los}).

The accuracy of the distances in the extinction cube is, however, limited by the fact that the stellar distances were derived by simple inversion of parallaxes, introducing significant bias to the  absolute distance scale. In order to eliminate this bias, ancillary data sets with Bayesian estimates of the distances and other stellar parameters have been created. The currently most comprehensive distance set is the \ga\ DR2 \sh\ data set \citep{2019A&A...628A..94A}. This catalogue contains 265 million stars with distance, extinction, and other parameters estimated by the \sh\  Bayesian code \citep{2018MNRAS.476.2556Q} using data from  \ga, Pan-STARRS1, 2MASS, and AllWISE. The catalogue is available at \url{gaia.aip.de}.

\begin{figure}
\resizebox{\hsize}{!}{\includegraphics[trim=0.2cm 0.1cm 0.cm 0.cm,clip=true,width=0.49\textwidth]{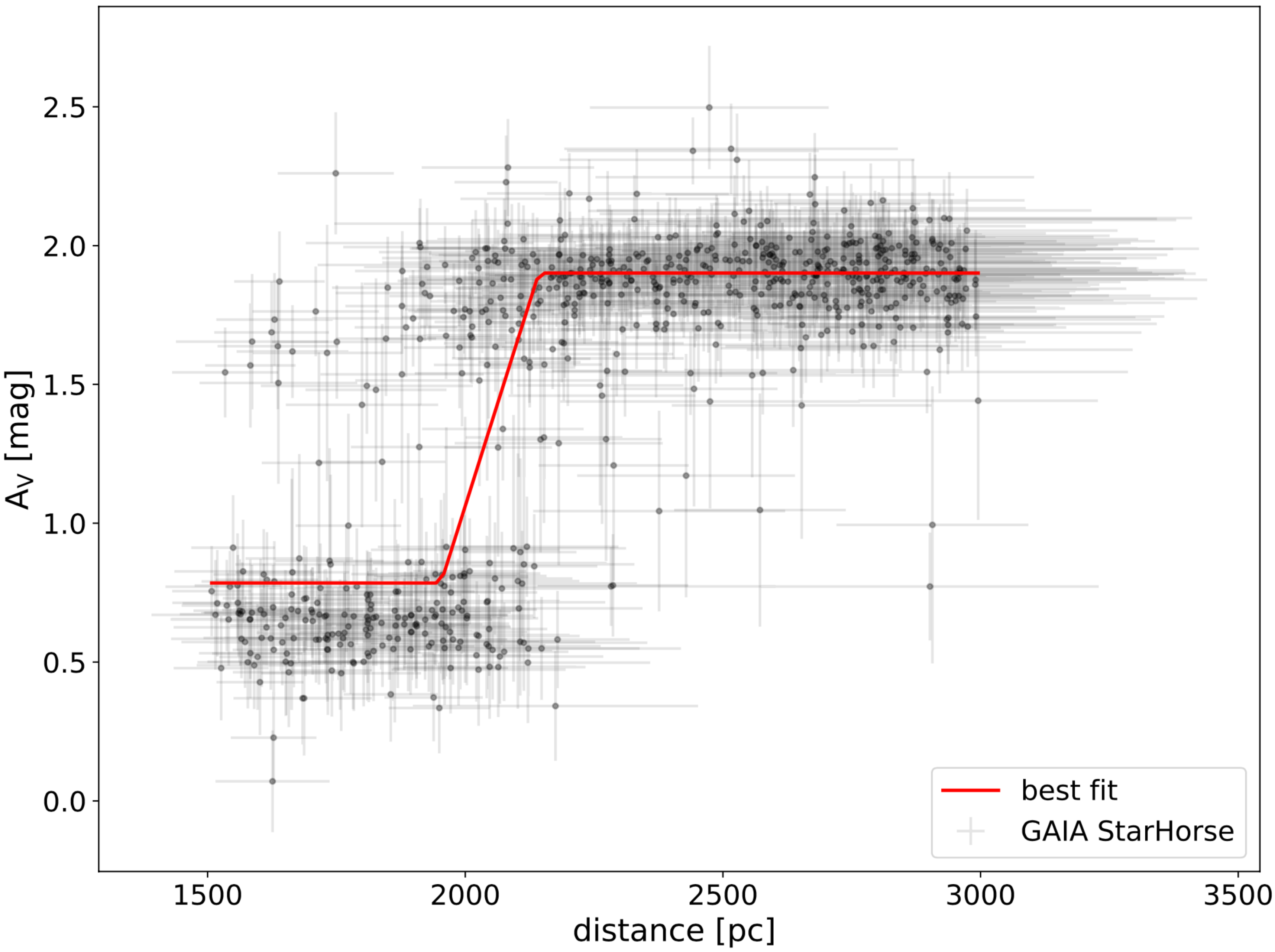}}
    \caption{Orthogonal distance regression fit of a homogeneous extinction layer model in the distance--$A_\mathrm{V}$ plane of the \sh\  data. The depth of the dust layer was fixed to 190\,pc.}
   \label{f:sh_fit}
\end{figure}

In order to determine the dust distribution, we utilise the subset of the catalogue for the region within 0\fdg9 of the position of \maxi, covering the whole extent of the scattering ring in eRASS1 and eRASS2. We then selected the objects in the area of ``red clump'' giant stars in the dereddened $M_0$--$(\mathrm{B}-\mathrm{R})$-plane for our analysis, since these stars are luminous enough to cover the relevant distances and their stellar parameters have accurate estimates. We again consider only stars with relative uncertainties in distance $< 20\%$. When plotting extinction versus distance for the resulting sample, a steep increase in $A_\mathrm{V}$ is visible at 2000\,pc. It is this region where the scattering ring is formed. In order to determine the distance of the dust causing this extinction, we fit a model for the $A_\mathrm{V}$ of a simple, homogeneous layer of dust to the data in the distance interval $1500-3000$\,pc (Fig.~\ref{f:sh_fit}). The free parameters of the fit are $A_{\mathrm{V},0}$ (on the near side of the layer), $A_{\mathrm{V},1}$ (behind the dust) and $d_0$ (distance). The geometrical depth of the dust sheet is well constrained by the X-ray 
images of the scattering ring and is fixed at 190\,pc (see Sect. \ref{s:distance} and Table \ref{tab:distances}). The best fitting mid-point distance of the dust sheet is at  $2047 \pm 22 \mathrm{pc}$, the best fit extinction values are $A_{\mathrm{V},0} = 0.78$\,mag, $A_{V,1} = 1.90$\,mag (see Fig.~\ref{f:sh_fit}). The resulting distance remains very stable even if the depth of the layer is left free to vary.

\subsection{Distance towards \maxi }
\label{s:distance}

\begin{table*}
  \caption{Results of distance measurements.}
  \label{tab:distances}
    {\centering
    \begin{tabular}{l@{}cccc}
      \hline\hline\noalign{\smallskip}
      Data set & $d_{\mathrm{dust}}$ & $xD$ & $x$ & $D$ \\
            &   pc          &    pc &     &  pc  \\
      \hline\noalign{\smallskip}
      eRASS1 &  $180^{+24}_{-23}$     & $2046^{+23}_{-23}\pm 205$      &  $0.6035^{+0.0046}_{-0.0036}$   & $3390 ^{+46}_{-43} \pm 339 $   \\[1ex]
      XMM EPIC\ \    &  $190^{+10}_{-8}$      & $2047^{+22}_{-22}\pm 205$       &  $0.6029^{+0.0018}_{-0.0012}$   & $3395^{+38}_{-37}\pm  340$     \\[1ex]

      \hline
    \end{tabular}}
    \tablefoot{\tablefoottext{*}\ For the parameters $xD$ (distance to the dust layer) and $D$ (distance to \maxi), the statistical uncertainty and the 10\%  systematic error due to \ga\ parallax uncertainties are listed. The error bars in $D$ are calculated by quadratically adding the statistical uncertainties for $xD$ and $x$.)}
\end{table*}

With the distance $xD$ well constrained by the \sh\ data, the distance $D$ to \maxi\ can be determined using the geometry shown in Fig.~\ref{f:sketch}. However, as discussed above, both the distribution of time lags resulting from the burst light curve (Fig.~\ref{f:maxi_lc}) and the distribution of dust along the line of sight contribute to the width and the radial profile of the ring. For an accurate determination of $D$ we therefore modelled the radial profile with the following steps:

1. As for the fit in Fig.~\ref{f:sh_fit} we assume a homogeneous dust layer of a certain thickness $d_{\mathrm{dust}}$ at a distance $xD$. The layer was divided into  slices of 1 pc depth. For each slice and angle $\theta$ we calculated the X-ray flux $F_\mathrm{X}(\theta)$ from the 2-3 keV MAXI lightcurve (Fig.~\ref{f:maxi_lc}) using the matching time delay $\Delta t$ at the time of the observation according to Eq.~\ref{eq:theta}.

\text
2. Following \citet{mathis_lee91}  and \citet{2011ApJ...738...78X} the observed brightness distribution 
as a function of $\theta$ from each dust slice at relative distance $x$ is given by
\begin{equation}
  \label{eq:xiang}
   I(\theta, E)   = F_\mathrm{X}(\theta,E) N_\mathrm{H} (1-x)^{-2} \;\frac{d\sigma (E,\theta_{sca})}{d\Omega}
\end{equation}
where $F_\mathrm{X}(E,\theta)$ is the source X-ray flux relevant at angle $\theta$ as determined in step 1, $N_\mathrm{H}$ is the  hydrogen column density in the distance slice, and $\theta_{sca}\sim\theta / (1-x)$.
The scattering cross section ${d\sigma}/{d\Omega}$ depends on the composition and size distribution of the
dust. \citet{draine03} gives easy-to-use analytical approximations to the cross sections for the dust model by  \citet{2001ApJ...548..296W} which we adopt here:
\begin{equation}
  \label{eq:draine}
   \frac{d\sigma}{d\Omega}   =  \frac{\sigma_{sca}}{\pi \theta^2_{s,50}} \frac{1}{\left[1+(\theta_{sca} / \theta_{s,50})^2\right]^2}
\end{equation}
We evaluated Eq. \ref{eq:draine} at $E=1.5\,\mathrm{keV}$ where the characteristic scattering angle is $\theta_{s,50}= 4'$ \citep{draine03}.
We made no attempt to model the absolute flux of the scattering ring, hence in Eqs.~\ref{eq:xiang} and \ref{eq:draine} only the dependencies on $\theta$ are important here. Since at a given epoch the scattering ring covers a relatively small range of  angles  $\theta$, the exact function of $d\sigma (\theta_{sca}) /d\Omega$ only marginally changes the model profile $I(\theta)$. Hence the choice of the model on dust composition and grain size distribution has only negligible influence on our results.

3. The ring profiles from each distance slice were added and the total flux  normalised to the observed flux. The result is a model profile for the parameter pair $d_{\mathrm{dust}}$ (depth of the dust layer) and $D$
(source distance). To find the best fitting values for the parameters $D$ and $d_\mathrm{dust}$, we calculated the profiles for a grid of $D$ and $d_\mathrm{dust}$ values and then determined the $\chi^2$-values for each resulting profile with respect to the measured profiles from eRASS1 and \xmmn. 
Since the X-ray images constrain the depth of the dust layer better than the \ga data, we re-iterated fitting $xD$ to the \sh\ data with the $d_\mathrm{dust}$ values derived from X-rays.
The parameter space covered in the final run was $140 \mathrm{pc} < d_\mathrm{dust} < 220 \mathrm{pc}$ ,
$3280 \mathrm{pc} < d_\mathrm{dust} < 3480 \mathrm{pc}$ for the eRASS1 profile and $175 \mathrm{pc} < d_\mathrm{dust} < 205 \mathrm{pc}$, $3350 \mathrm{pc} < d_\mathrm{dust} < 3440 \mathrm{pc}$ for the \xmmn profile.

\begin{figure*}
\resizebox{0.499\hsize}{!}{\includegraphics[trim=1.5cm 1cm 1cm 1cm,clip=true]
{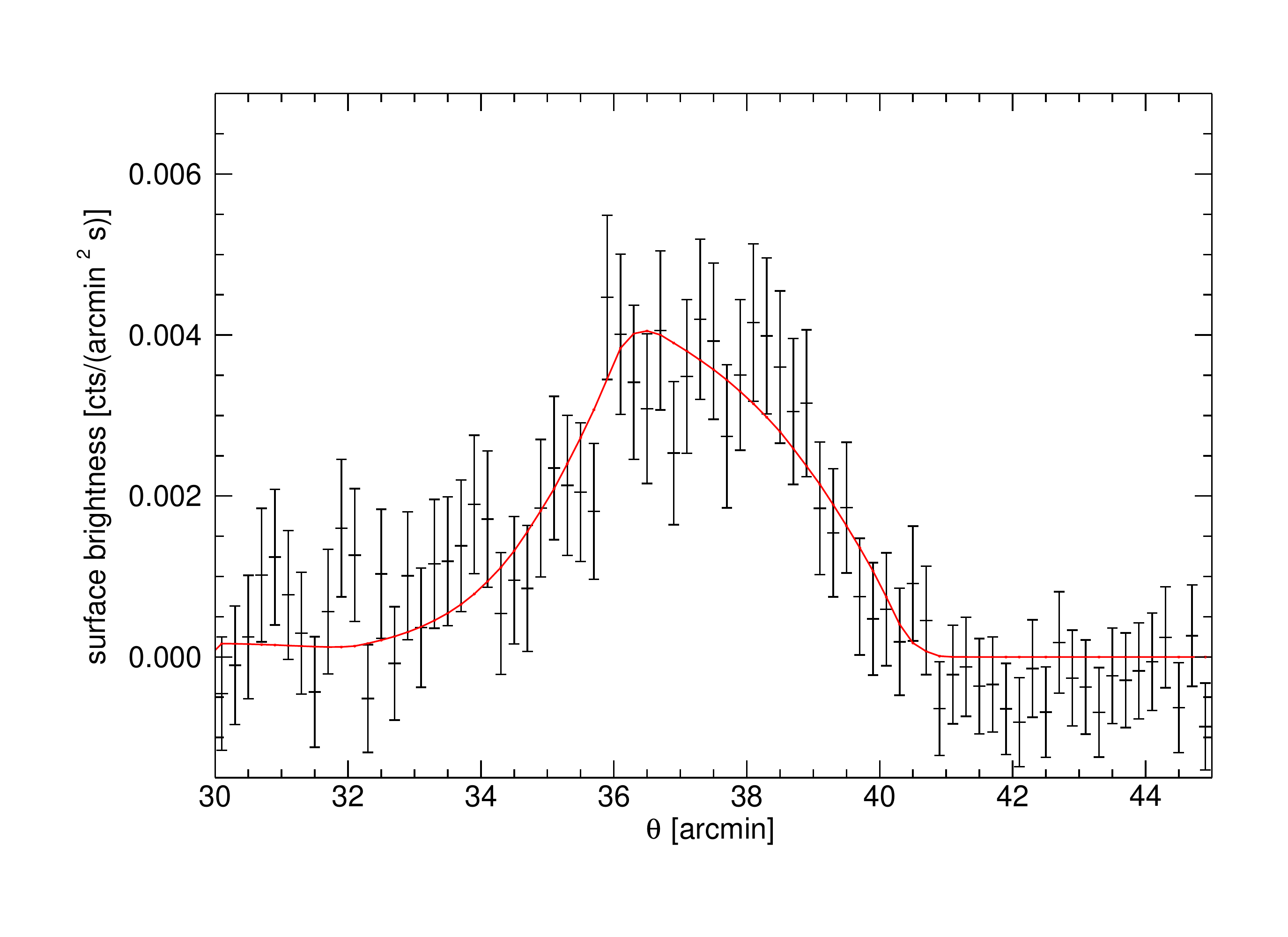}} \hfill
\resizebox{0.499\hsize}{!}{\includegraphics[trim=1.5cm 1cm 1cm 1cm,clip=true]
{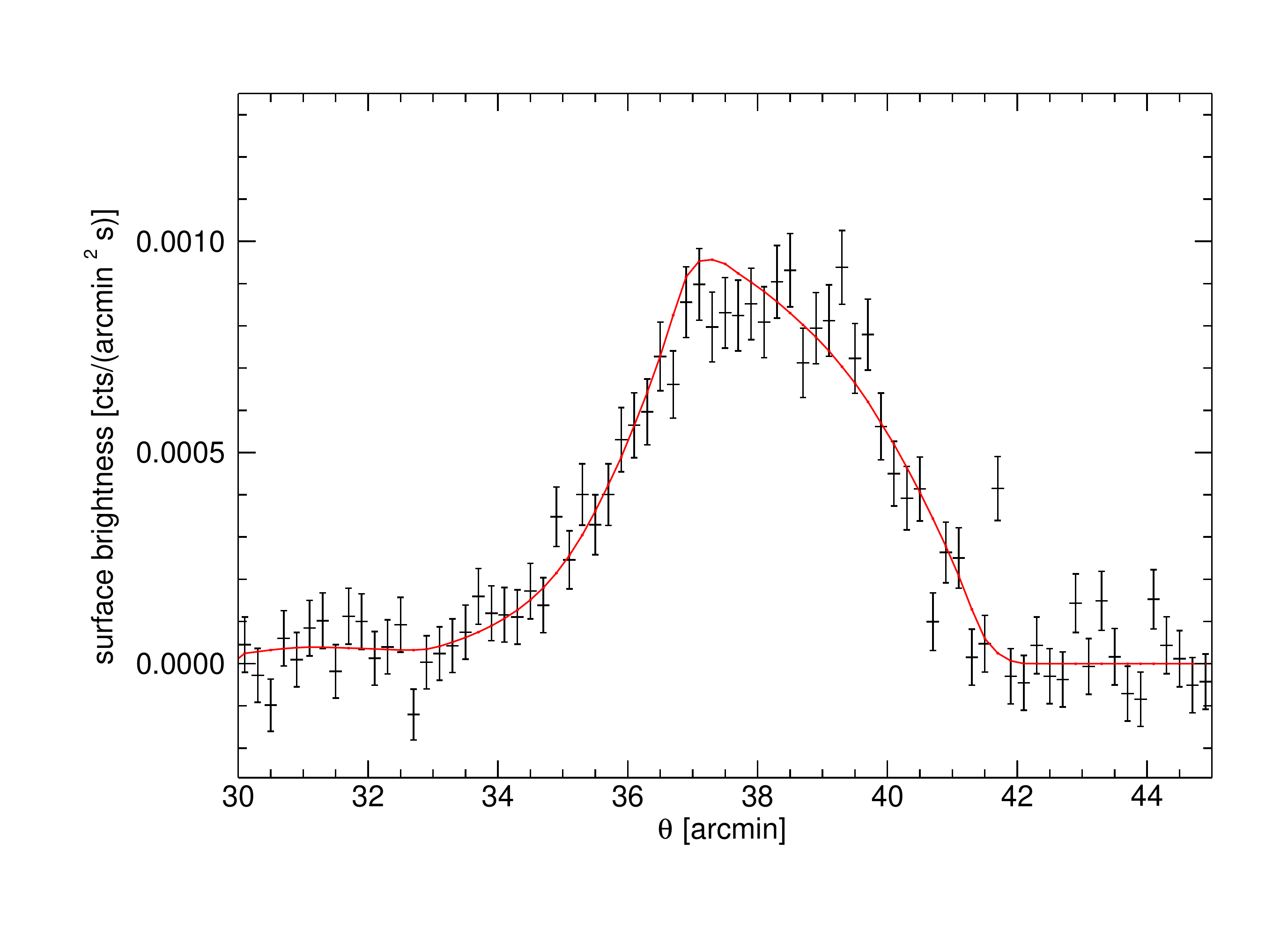}}
    \caption{{\it (left)} Background subtracted eRASS1 ring profile (1.0-2.3 keV) with best fit model (red line , $D=3390\,\mathrm{pc}$, $d_\mathrm{dust}=180\,\mathrm{pc}$).
   {\it (right)} Background subtracted XMM EPIC ring profile (1.0-2.0 keV) with best fit model (red line , $D=3395$\,pc, $d_{\rm dust}=190$\,pc).
    }
   \label{f:erass1_fit}
\end{figure*}

The fit results for both the eRASS1 and \xmmn\ observations are presented in Table~\ref{tab:distances} and
Fig. \ref{f:erass1_fit}.
It should be noted that using the dust distribution given by the extinction cube by \citet{2019A&A...625A.135L} (Fig.~\ref{f:dust_los}) in the model results in a ring profile which is much broader and more structured than the observed profile, hence we conclude that our model assumption of a simple homogeneous layer is a better approximation to the actual distance distribution.

The distance ratio $x$ between the scattering dust and \maxi can be measured with remarkable precision: 0.25\% for the \xmmn\ observation and 0.7\% for eRASS1. For the \xmmn\ data the total statistical uncertainties (1.1\%) are dominated by the errors in the distance towards the dust layer.

The absolute accuracy of the distance essentially depends on the systematic parallax uncertainties in the \ga\ DR2 catalogue, which are still under investigation. Analysis of QSO parallaxes revealed a global negative parallax zero-point but also  spatial variations \citep{2018A&A...616A...2L}. A systematic uncertainty of ${\sim} 0.05\,\mathrm{mas}$ has to been taken into account in the absolute accuracy  of distances. In our case this amounts to a 10\% absolute accuracy of the measurement to the scattering dust at 2\,kpc, and hence the same relative accuracy for the distance towards \maxi. The same level of uncertainty has also been adopted by \citet{2020A&A...633A..51Z} for \ga\ based distances of molecular clouds in the 2\,kpc range. Since the statistical precision of the measurement with respect to the \ga\ distance frame is much better, however, improvements of the absolute \ga\ distances in future data releases will directly lead to a higher accuracy of the \maxi\ distance.

\subsection{X-ray spectral analysis of the dust scattering ring}
\label{s:ringspec}

\begin{figure}
\includegraphics[width=\linewidth]{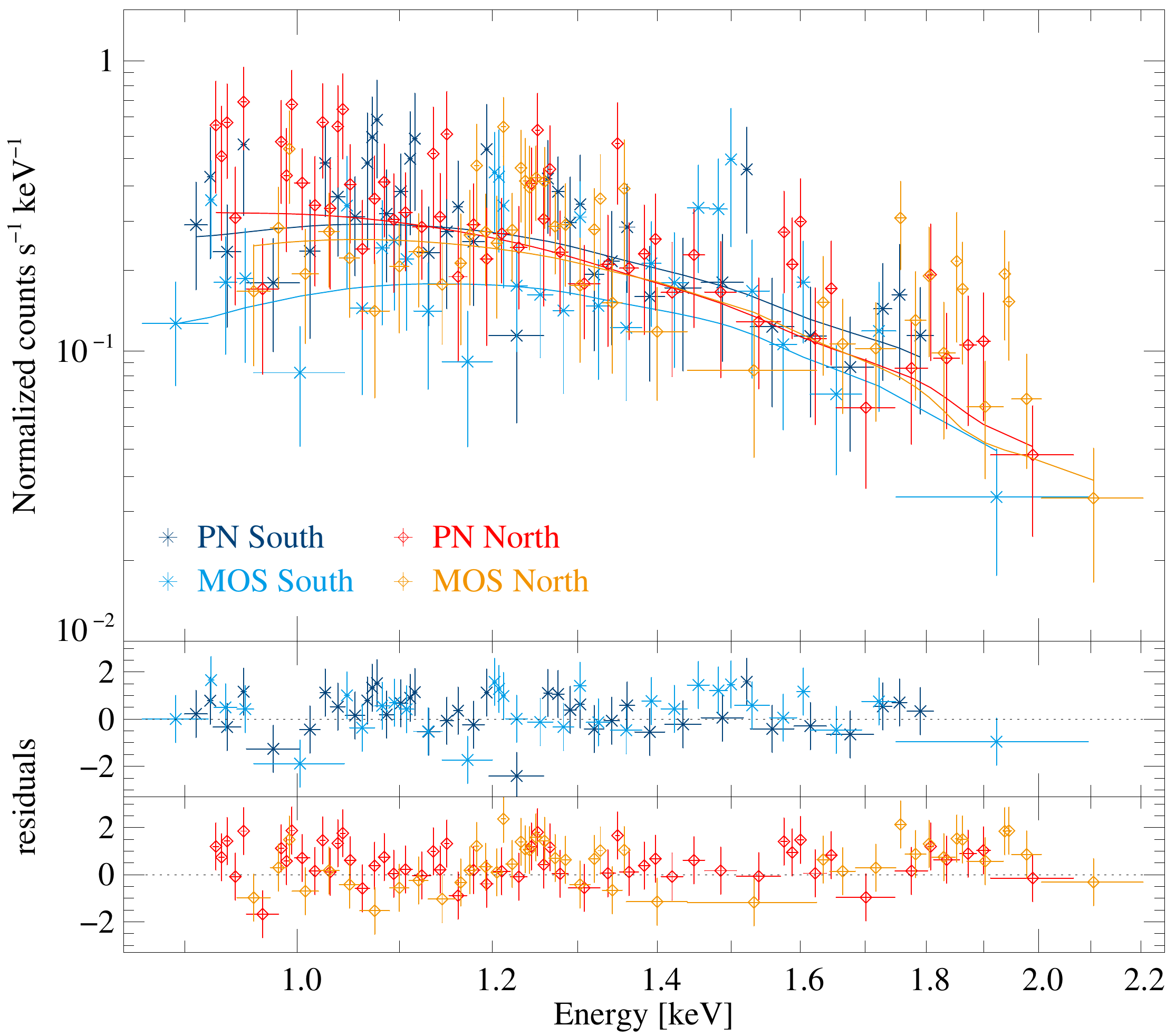}
\caption{\textit{XMM-Newton} ring spectra taken by EPIC/pn and MOS in the Northern pointing 2 (blue crosses) and the Southern pointing 6 (red diamonds) with an absorbed power-law model. The data have been binned to a signal-to-noise ratio of at least 2 for clarity. The residuals (data$-$model)/error are shown separately in the lower panels.}
     \label{fig:epicspectra}
\end{figure}

\begin{table}
  \caption{Parameters of the fits to \textit{XMM-Newton} EPIC ring
    spectra with absorbed power-law models\tablefootmark{*}, all performed with $\chi^2$ fit statistics}
  \label{tab:epicspectra}
    {\centering
    \begin{tabular}{l@{}c@{~~}c@{~~~}c@{~~~}c}
      \hline\hline\noalign{\smallskip}
      Parameter & Units & Model 1 & Model 2 & Model 3\\
      \hline\noalign{\smallskip}
      $N_\mathrm{H}$  &  $10^{21}\,\textrm{cm}^{-2}$
        & $8.0^{+1.2}_{-1.1}$ & $7.5^{+0.4}_{-0.4}$ &  \\[1ex]
      $N_{\mathrm{H,North}}$  &  $10^{21}\,\textrm{cm}^{-2}$
        &                   &                   & $8.7^{+0.7}_{-0.7}$ \\[1ex]
      $N_{\mathrm{H,South}}$  &  $10^{21}\,\textrm{cm}^{-2}$
        &                   &                   & $6.8^{+0.5}_{-0.5}$ \\[1ex]
      $\Gamma$ & 
        & $4.2^{+0.4}_{-0.4}$       & 4.0               & 4.0               \\[1ex]
      norm  &  
        & $1.7^{+0.5}_{-0.4}$ & $1.5^{+0.2}_{-0.2}$ & $1.4^{+0.2}_{-0.2}$ \\[1ex]
       \multicolumn{3}{c}{($ 10^{-5}\,\mathrm{ph}\,\textrm{keV}^{-1}\,\textrm{cm}^{-2}\,\textrm{arcmin}^{-2}$)}  & &                                                                 \\[1ex]
        $\chi^2_\textrm{red} (dof)$ & & 0.9 (1258) & 0.9 (1259) & 0.9 (1257) \\[1ex]
      \hline
    \end{tabular}}
    \tablefoot{\tablefoottext{*}\ Model \texttt{const*tbabs(powerlaw), abundance=wilm} in
      \xspec. Model 1: $N_\mathrm{H}$ and power-law parameters allowed to vary. Model 2: power-law index fixed to 4.0. Model 3: Separate absorption terms for the three Northern and the three Southern pointings. All parameter errors correspond to 90$\%$ confidence limits
for 1 parameter.}
\end{table}

The \textit{XMM-Newton} EPIC spectra of the dust ring were binned to reach signal-to-noise ratios of at least 1.0 per bin in the energy range between 0.7\,keV and 2.2\,keV in the individual pointings. We fitted them jointly with an absorbed power-law, using the absorption model and abundances of \citet{2000ApJ...542..914W}. A multiplicative factor accounted for the different extraction regions and fluxes of the spectra (\xspec model \texttt{const*tbabs(powerlaw)}). The best fit ($\chi^2_\textrm{red}=0.9$ for 1\,258 degrees of freedom) resulted in 
$N_\textrm{H}=8.0^{+1.2}_{-1.1}\times 10^{21}\,\textrm{cm}^{-2}$ and a power-law index of $\Gamma=4.2\pm0.4$. This $N_\textrm{H}$  is  consistent with the values obtained for \maxi from other missions.
\citet{2020ApJ...899L..20T} modelled the MAXI spectrum during the high/soft state  with a disk black body model and upscattering at higher energies  (\texttt{tbabs *simpl * diskbb}). The resulting disk temperatures at the innermost radius $kT_{Rin}$ are in the range 0.6-0.75 keV. In the energy range 
contributing to the dust scattering ring such a disk black body can be approximated by a power-law with $\Gamma \sim 2$.  When taking into account the modification of the incident spectrum by the energy dependence of the scattering cross section which is approximately  $E^{-2}$, an absorbed power law with $\Gamma \sim 4$ can be expected for the ring spectrum.  Fixing the power-law index at the  $\Gamma=4.0$, the absorption is better constrained to $7.5\pm0.4\times 10^{21}\,\textrm{cm}^{-2}$. To investigate the spatial dependence of the absorption terms, we employ two independent values for the Northern and for the Southern part of the dust ring, coupling pointings 2, 10, 11 in the North and 6, 7, 8 in the South. For the fixed power-law index of 4.0, we measure $N_\textrm{H,North}=8.7\pm0.7\times 10^{21}\,\textrm{cm}^{-2}$ in the Northern part of the ring and $N_\textrm{H,South}=6.8\pm0.5\times 10^{21}\,\textrm{cm}^{-2}$ in the Southern part. Figure~\ref{fig:epicspectra} shows example spectra of a Northern and a Southern pointing and Table~\ref{tab:epicspectra} the full list of model parameters.

For both the eRASS1 and eRASS2 observations, event files from the latest eSASS  pipeline  version (c946) were used to extract spectra of the ring area using the \texttt{srctool} task. For the eRASS1 data an annulus around the position of \maxi\ with radii $R_1=34.4$ arcmin, $R_2=40.6$ arcmin was used to extract the source events. For eRASS2 events in the annulus between $R_1=41.8$ arcmin, $R_2=49.3$ arcmin were extracted.
In both cases the background was extracted from annuli larger than the source annuli, intervening sources were excised from the source and background regions.
Given the limited SNR in the eRASS2 spectrum, we do not expect to measure a spectral index or $N_\mathrm{H}$  significantly different from eRASS1. Therefore we only fit this spectrum together with the eRASS1 spectrum and tie the  $\Gamma$ and $N_\mathrm{H}$ parameters,
leaving only the normalisations free to vary independently.
The resulting model parameters  are compiled in Table~\ref{tab:erospec}.
The spectral indices and absorbing columns densities are in line  with the \xmmn measurements. When fixing $N_\mathrm{H}$ at the value measured for \maxi \citep{2020ApJ...899L..20T},
$8.6 \times 10^{21}\,\textrm{cm}^{-2}$, the best fit photon index is ${\sim} 4.0$,
as expected for an incident spectrum with $\Gamma \sim 2$.
When fitting  both the eRASS1 and eRASS2 spectra with  fixed $N_\mathrm{H}=8.6 \times 10^{21}\,\textrm{cm}^{-2}$  and $\Gamma=4.0$, the 
resulting 0.5--2.0\,keV fluxes are $5.32\pm0.27 \times 10^{-12}\, \mathrm{erg}\,\mathrm{cm}^{-2}\,\mathrm{s}^{-1}$
(eRASS1) and $1.57\pm0.27 \times 10^{-12}\, \mathrm{erg}\,\mathrm{cm}^{-2}\,\mathrm{s}^{-1}$.
With $3.4 \pm 0.61 (1 \sigma)$ the eRASS1 $/$ eRASS2 flux ratio is somewhat higher than the factor $~(1.2)^{4}=2.1$ expected due to the increasing scattering angles and the $\theta^{-4}$ function of the  \citet{draine03} cross sections.
This might be an indication for a decrease of the scattering cross sections steeper than $\sim \theta^{-4}$. On the other hand one has to consider the large linear size of the ring with a diameter of $\sim 50$ pc during the time of the eRASS2 observations. Given the azimuthal brightness variations visible in Fig. \ref{fig:epicmosaic}, variations in the dust distribution may also contribute to the observed flux ratio.

 \begin{table}
\caption{Parameters of the spectral fits to the dust scattering rings in eRASS1 and eRASS2.}
\label{tab:erospec}
{\centering
    \begin{tabular}{l@{}cccc}
      \hline\hline\noalign{\smallskip}
      Param. & Units & Model1 & Model2 & Model3\\
      \hline\noalign{\smallskip}
      \multicolumn{5}{c}{eRASS1} \\
      \hline\noalign{\smallskip}
      $N_\mathrm{H}$  &  $10^{21}\,\textrm{cm}^{-2}$
        & $11.1^{+3.5}_{-2.8}$        & $8.6$                   &   $8.4 ^{+0.9}_{-0.8}$    \\[1ex]
      $\Gamma$ & 
        & $5.04^{+1.25}_{-1.02}$      & $4.17^{+0.30}_{-0.31}$ &  4.0  \\[1ex]
      norm &  $10^{-3}\,\mathrm{ph}\,\textrm{keV}^{-1}\,\textrm{cm}^{-2}$
        & $23.0^{+21.6}_{-9.7}$ & $14.2^{+1.3}_{-1.4}$ & $13.3^{+2.0}_{-1.8}$ \\[1ex]
       $\chi^2_\textrm{red} (dof)$ & & 1.0 (284) & 1.0 (285) & 1.0 (285) \\[1ex] 
      \hline\noalign{\smallskip}
      \multicolumn{5}{c}{eRASS1 + eRASS2} \\
      \hline\noalign{\smallskip}
      $N_\mathrm{H}$  &  $10^{21}\,\textrm{cm}^{-2}$
        & $12.6^{+4.0}_{-3.0}$         & $8.6$           &  $8.7 ^{+0.9}_{-0.8}$    \\[1ex]
      $\Gamma$ & 
        & $5.42^{+1.32}_{-1.1.06}$  & $4.09^{+0.29}_{-0.29}$ &  4.0  \\[1ex]
      norm1 &  $10^{-3}\,\mathrm{ph}\,\textrm{keV}^{-1}\,\textrm{cm}^{-2}$
        & $29.06^{+31.0}_{-13.1}$ & $14.0^{+1.3}_{-1.4}$ & $14.0^{+2.0}_{-1.8}$ \\[1ex]
      norm2 &  $10^{-3}\,\mathrm{ph}\,\textrm{keV}^{-1}\,\textrm{cm}^{-2}$
        & $9.1^{+10.7}_{-4.5}$   & $4.1^{+1.2}_{-1.2}$  & $4.1^{+1.4}_{-1.2}$ \\[1ex]
        $\chi^2_\textrm{red} (dof)$ & & 1.0 (570) & 1.0 (571) & 1.0 (571) \\[1ex]
      \hline
    \end{tabular}}
    \tablefoot{\tablefoottext{*}\ Model \texttt{const*tbabs(powerlaw), abundance=wilm} in
      \xspec. Model 1: $N_\mathrm{H}$ and power-law parameters allowed to vary. Model 2: $N_\mathrm{H}$  fixed to $8.6 \times 10^{21}\,\textrm{cm}^{-2}$. Model 3: power-law index fixed to 4.0. 
      For the eRASS1 + eRASS2 fits both spectra were fitted simultaneously with $N_\mathrm{H}$ and  $\Gamma$ tied between the spectra and norm1 and norm2 left to vary independently. All parameter errors correspond to 90$\%$ confidence limits
for 1 parameter.}
\end{table}

\subsection{X-ray spectral analysis of \maxi in the  post-outburst phase}

\subsubsection{eRASS1 \& eRASS2}

In  the eRASS1 image \maxi\ was clearly detected, in a circular area with 0.5 arcmin radius we extracted 95 net counts (0.18 cts/s). The spectrum was fitted in \xspec with an absorbed power law (\texttt{tbabs*powl}). Since the absorbing column density is poorly constrained, it was again fixed to $N_\textrm{H}=8.6 \times 10^{21}\,\textrm{cm}^{-2}$. The best fit spectral index in this case is $\Gamma = 1.51 ^{+0.46}_{-0.45}$, the resulting 0.5 - 2.0 keV flux is $(3.04\pm0.85) \times 10^{-13}$\,erg cm$^{-2}$ s$^{-1}$. During the eRASS2 observation \maxi  was not detected, the upper limit flux between 0.5-2.0 keV is $3 \times 10^{-14}$
\,erg cm$^{-2}$ s$^{-1}$.

\subsubsection{\xmmn}

MAXI\,J1348$-$630 was faint at the time of the \textit{XMM-Newton} observation with a mean count rate of about $0.005\,\mathrm{cts}\,\mathrm{s}^{-1}$. The EPIC spectra were fitted with an absorbed power law and W statistic (\texttt{cstat} in \xspec), resulting in a poorly constrained
$N_\textrm{H}=7.7^{+7.8}_{-4.9}\times 10^{21}\,\textrm{cm}^{-2}$
and a power-law index of
$\Gamma=1.9^{+0.8}_{-0.7}$.
From the fit to the EPIC/pn spectrum, we derive an absorbed 0.5--2.0\,keV flux of $1.1^{+0.1}_{-0.3}\times 10^{-14}\,\textrm{erg\,cm}^{-2}\,\textrm{s}^{-1}$, roughly a factor 10 lower than during eRASS1 just three weeks before the \xmmn\ data were obtained. Fixing column density and power-law index at the eRASS1-derived values, we obtain a flux of $9.2^{+1.3}_{-1.6}\times 10^{-15}\,\textrm{erg\,cm}^{-2}\,\textrm{s}^{-1}$ in the same band.

\section{Discussion and Conclusion}\label{sec:discuss}

Since the first idea about dust scattering halos was formulated and after the first successful attempt of its realisation almost three decades later, X-ray scattering on interstellar dust has become an established method of determining distances. The method works particularly well when the X-ray source shows bursts, because then instead of a uniform halo, expanding rings can usually be observed. The ambiguity between the distance of the source and the position of the dust can be resolved if one of the two quantities is known. If it can be assumed to be infinite, as in the case of GRBs, the distance between the interstellar dust layers can be determined.

\maxi\ presents an ideal case for such studies, since the only layer of dust in between is already precisely known based on \ga\ measurements, so that we were able to measure the distance of the X-ray source itself. The distance of 3390\,pc has very low statistical uncertainty (1.1\%) with respect to the \ga\ distance frame, the absolute accuracy is limited by systematics of the \ga\ parallaxes which amount to an uncertainty of 10\%. This makes the distance to \maxi\ one of the best known distances to a black hole binary, which were typically measured with Very Long Baseline Interferometry with typical statistical uncertainties on the order of 5--10\%. Examples include Cyg X-1, with $1.86^{+0.12}_{-0.11}$\,kpc \citep{2011ApJ...742...83R}, the systematic uncertainty of this position is large, given its \ga\ DR distance of $2.39^{+0.21}_{-0.18}$\,kpc \citep{2019MNRAS.485.2642G}, which makes it consistent with newer VLBI data (Miller-Jones et al., 2020, submitted). A distance to Cyg~X-1 has also been calculated by the analysis of its dust scattering halo by \citet{2011ApJ...738...78X}, who arrive at a distance in the interval $1.72- 1.90$ kpc.

Our distance improves the earlier distance estimates for \maxi, which were based on not well-calibrated distance indicators such as the transition luminosity between the hard and the soft state. Assuming that this luminosity is between 1\% and 4\% of the Eddington luminosity \citep[][and references therein]{1995PASP..107.1207N,2003A&A...409..697M,2019MNRAS.485.2744V}, 
\citet{2020ApJ...897....3J} estimate the distance of \maxi\ to 5--10\,kpc, consistent with the distance estimate based on the dereddened, $A_\mathrm{V}= 2.4$, optical magnitude and the position of the object in the optical-X-ray luminosity diagram, which yields $D=3$--8\,kpc \citep{2019ATel12439....1R}. Using the same ansatz, \citet{2020ApJ...899L..20T} estimate the distance to \maxi\ to 4--8\,kpc. Noting that with $N_\mathrm{H} \sim 8.6 \times 10^{21}\,\mathrm{cm}^{-2}$
the measured X-ray absorption column is significantly lower than the integrated 21\,cm interstellar column density     \citep[$N_\mathrm{H}\sim (1.45\ldots1.53)\times 10^{22}\,\mathrm{cm}^{-2}$,][]{2016A&A...594A.116H}, however, \citeauthor{2020ApJ...899L..20T} argue that \maxi\ must be in front of the Scutum-Centaurus arm (consistent with our result), and then argue for a most likely distance of 3.8\,kpc.

Using the MAXI monitoring, \citet{2020ApJ...899L..20T} utilised the measured evolution of the inner radius of the accretion disk during the soft state to estimate the mass of the black hole \citep[e.g.,][and references therein]{2010ApJ...718L.117S}. With our improved distance, we can revise their inner disk radius measurements to $R_\mathrm{in}=97\pm 13\,\mathrm{km}$ (including a systematic error of 10\%), leading to a revised estimate of the black hole mass of
\begin{equation}
M_\mathrm{BH}= \frac{c^2 R_\mathrm{in}}{6G} \sim
(11 \pm 2)\, \left(\frac{D}{3.39\,\mathrm{kpc}}\right) (\cos i)^{-1/2}\,M_\odot
\end{equation}
compared to the earlier estimate of $13\pm2\,M_\odot$ \citep{2020ApJ...899L..20T}. The transition to the soft state, which MAXI measured at a bolometric flux of $\sim1.7\times10^{-8}\,\mathrm{erg}\,\mathrm{s}^{-1}\,\mathrm{cm}^{-2}$ therefore occurred at a luminosity of 1.7\% of the Eddington luminosity.
With our revised distance and mass the peak flux of $1.0\times10^{-7}\,\mathrm{erg}\,\mathrm{s}^{-1}\,\mathrm{cm}^{-2}$ given by \citet{2020ApJ...899L..20T}
corresponds to $10\%$ of the Eddington luminosity.

\begin{figure}
\resizebox{\hsize}{!}{\includegraphics{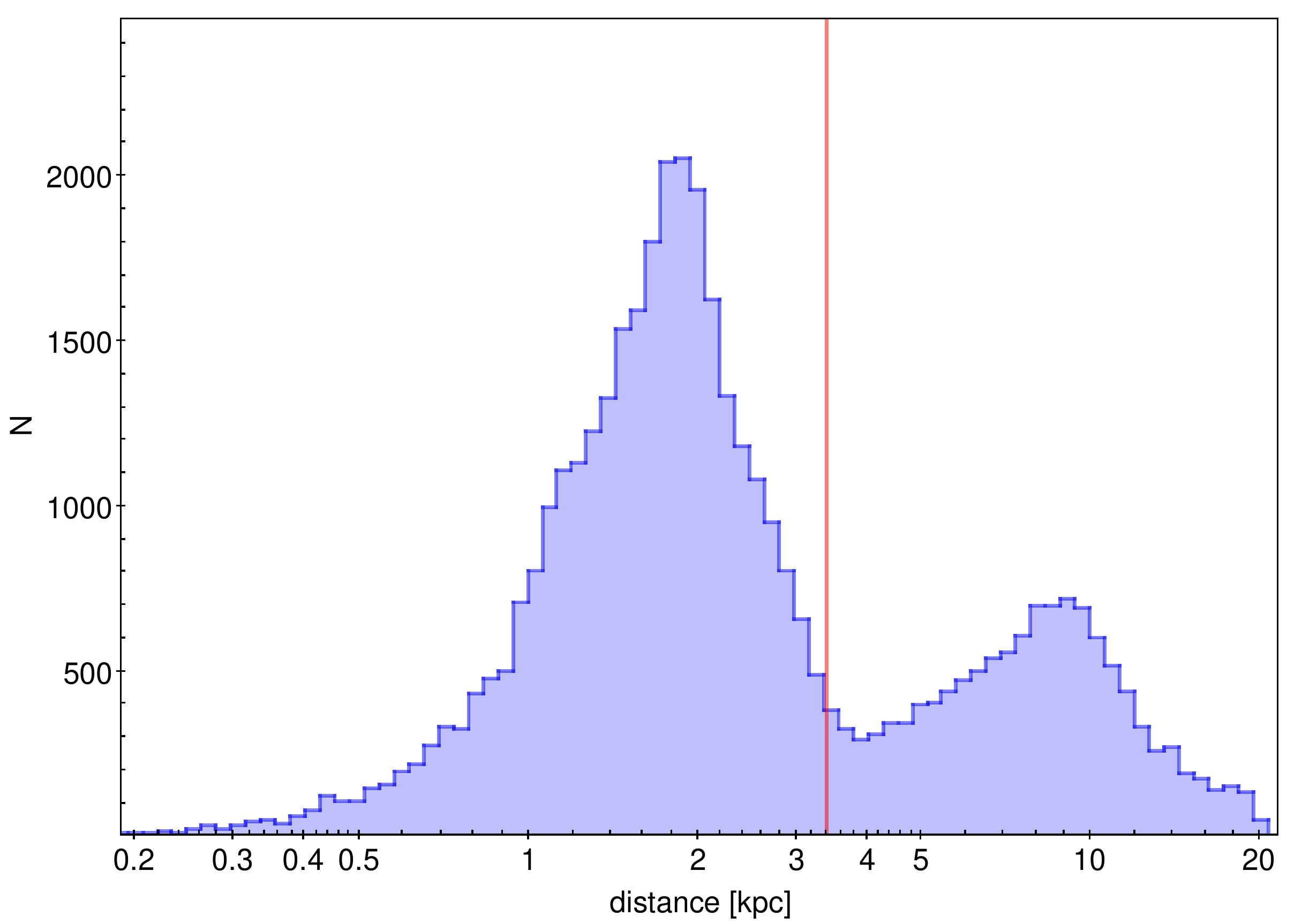}}
    \caption{Distance of \maxi\ (red) compared to the distribution of distances in the \ga\ \sh\ data  within 0.9 deg from the LOS.}
   \label{f:sh_histo}
\end{figure}

The distance measurement places \maxi\ at a position between the Sagittarius and Scutum-Centaurus spiral arms of the Milky Way. Figure~\ref{f:sh_histo} shows that \maxi is located in an area of relatively low stellar density. With our distance modulus $\mu= 12$, $A_V=2.4$ and the quiescence brightness $r'=20.69$  \citep{baglio+20} we derive an absolute magnitude $M_{\rm r}=6.24$, consistent with a main sequence K-type donor star. Both the late type secondary and the location of the object suggest a relatively old age of the system.

The discovery of this giant scattering ring demonstrates the power of the \ero surveys  for this type of science, with their unlimited field of view and 6-monthly observing cadence. With the incidence of new black hole transients about once per year and the outburst activity of other powerful X-ray sources in our galaxy, we expect further discoveries of dust scattering halos and rings during the 4\,years of the survey and  significant insights  into the physics of both the transient sources and the interstellar medium.

\begin{acknowledgements}

We would like to thank the referee for useful suggestions which helped to improve the presentation of this paper.

We thank Friedrich Anders for the  useful discussion on the \ga\ \sh\ data set and the absolute accuracy of the \ga\ parallaxes.

This work is based on data from eROSITA, the primary instrument aboard SRG, a joint Russian-German science mission supported by the Russian Space Agency (Roskosmos), in the interests of the Russian Academy of Sciences represented by its Space Research Institute (IKI), and the Deutsches Zentrum für Luft- und Raumfahrt (DLR). The SRG spacecraft was built by Lavochkin Association (NPOL) and its subcontractors, and is operated by NPOL with support from the Max Planck Institute for Extraterrestrial Physics (MPE).

The development and construction of the eROSITA X-ray instrument was led by MPE, with contributions from the Dr.~Karl Remeis-Observatory Bamberg \& ECAP (FAU Erlangen-Nuernberg), the University of Hamburg Observatory, the Leibniz Institute for Astrophysics Potsdam (AIP), and the Institute for Astronomy and Astrophysics of the University of Tübingen, with the support of DLR and the Max Planck Society. The Argelander Institute for Astronomy of the University of Bonn and the Ludwig Maximilians Universität München also participated in the science preparation for eROSITA

The eROSITA data shown here were processed using the eSASS/NRTA software system developed by the German eROSITA consortium.

This work was supported by the Bundesministerium f\"ur Forschung und Technologie under Deutsches Zentrum f\"ur Luft- und Raumfahrt grants  50\,QR\,1603, 50\,QR\,1614, 50 OX 1901 and 50\,OX\,9562.

We thank the project scientist of \xmmn, Dr.~Norbert Schartel, for the generous allocation of observation time in Directors Discretionary Time.

This work has made use of data from the European Space Agency (ESA) mission
{\it Gaia} (\url{https://www.cosmos.esa.int/gaia}), processed by the {\it Gaia}
Data Processing and Analysis Consortium (DPAC,
\url{https://www.cosmos.esa.int/web/gaia/dpac/consortium}). Funding for the DPAC
has been provided by national institutions, in particular the institutions
participating in the {\it Gaia} Multilateral Agreement.
This research has made use of MAXI data provided by RIKEN, JAXA and the MAXI team.

\end{acknowledgements}

\bibliographystyle{aa}
\bibliography{dr}

\end{document}